\documentclass[useAMS,usenatbib]{mn2e}
\usepackage{graphicx}
\usepackage{times}
\usepackage{psfig}
\usepackage{amssymb}
\usepackage{latexsym}
\usepackage{natbib}

\title[NGC~4698] {Polar bulges and polar nuclear discs: the case of
  NGC 4698}

\author[E. M. Corsini et al.]{E. M. Corsini,$^{1,2,}$\thanks{E-mail:
    enricomaria.corsini@unipd.it} J. M\'endez-Abreu,$^{3,4}$
  N. Pastorello,$^{1,5}$ E. Dalla Bont\`a,$^{1,2}$ \and 
  L. Morelli,$^{1,2}$ A. Beifiori,$^{6}$ A. Pizzella$^{1,2}$ and
  F. Bertola$^{1,2}$\\
\\$^{1}$Dipartimento di Fisica e Astronomia `G. Galilei',
Universit\`a di Padova, vicolo dell'Osservatorio 3, I-35122 Padova,
Italy
\\$^{2}$INAF--Osservatorio Astronomico di Padova, vicolo
dell'Osservatorio 5, I-35122 Padova, Italy
\\$^{3}$Instituto Astrof\'{\i}sico de Canarias, Calle V\'{\i}a
L\'actea s/n, E-38200 La Laguna, Spain
\\$^{4}$Departamento de Astrof\'{\i}sica, Universidad de La Laguna,
Calle Astrof\'{\i}sico Francisco S\'anchez, E-38205 La Laguna, Spain
\\$^{5}$ Centre for Astrophysics and Supercomputing, Swinburne 
University of Technology, Hawthorn, VIC 3122, Australia
\\$^{6}$Max-Planck-Institut f\"ur extraterrestrische Physik,
Giessenbachstra{\ss}e, D-85748 Garching bei M\"unchen, Germany.}

\begin{document}

\date{Accepted 2012 March 20. Received 2012 March 19; in original form
  2011 December 23}

\pagerange{L\pageref{firstpage}--L\pageref{lastpage}} \pubyear{2012}

\maketitle

\label{firstpage}

\begin{abstract}

\noindent
The early-type spiral NGC~4698 is known to host a nuclear disc of gas
and stars which is rotating perpendicularly with respect to the galaxy
main disc. In addition, the bulge and main disc are characterised by a
remarkable geometrical decoupling. Indeed they appear elongated
orthogonally to each other.
In this work the complex structure of the galaxy is investigated by a
detailed photometric decomposition of optical and near-infrared
images.
The intrinsic shape of the bulge was constrained from its apparent
ellipticity, its twist angle with respect to the major axis of the
main disc, and the inclination of the main disc. The bulge is actually
elongated perpendicular to the main disc and it is equally likely to
be triaxial or axisymmetric.
The central surface brightness, scalelength, inclination, and
position angle of the nuclear disc were derived by assuming it is
infinitesimally thin and exponential. Its size, orientation, and
location do not depend on the observed passband.
These findings support a scenario in which the nuclear disc is the end
result of the acquisition of external gas by the pre-existing triaxial
bulge on the principal plane perpendicular to its shortest axis and
perpendicular to the galaxy main disc. The subsequent star formation
either occurred homogeneously all over the extension of the nuclear
disc or through an inside-out process that ended more than 5 Gyr ago.

\end{abstract}


\section{Introduction}

Several mechanisms have been proposed for bulge assembly. Bulges
formed either before the disc by hierarchical merging, or at the same
time in a monolithic collapse, or after the disc as a results of
secular evolution. Furthermore, there is evidence that bulges can
experience accretion events via infall of external material or
satellite galaxies.
None of these scenarios alone is able to reproduce the observed
properties of all the bulges along the Hubble sequence since bulges
are a class of diverse and heterogeneous objects. The large bulges of
lenticulars and early-type spirals are similar to low-luminosity
elliptical galaxies, whereas the small bulges of late-type spirals are
reminiscent of disc components (see \citealt{Kormendy2004} and
references therein).

The study of the intrinsic shape of bulges is a key constraint on
their formation.
Although the kinematics of many bulges is well described by dynamical
models of oblate ellipsoids which are flattened by rotation with
little or no anisotropy \citep[e.g.,][]{Pignatelli2001,
  Cappellari2006}, the twisting of the bulge isophotes
\citep{Lindblad1956, Zaritsky1986}, the presence of non-circular gas
motions \citep[e.g.,][]{Coccato2004, FalconBarroso2006}, and the
misalignment between the major axes of the bulge and disc
\citep[][hereafter MA+08]{Bertola1991, MendezAbreu2008} observed in
several galaxies cannot be explained if the bulge and disc are both
axisymmetric.  These features are interpreted as the signature of
bulge triaxiality.
Perfect axisymmetry is ruled out when the intrinsic shape of bulges is
determined by statistical analyses based on their observed ellipticity
(\citealt{Fathi2003}, MA+08).
A large fraction of bulges are characterised by an elliptical
equatorial cross-section, and most of them are flattened along the
polar axis they share with the surrounding disc. As a consequence,
triaxial bulges elongated perpendicularly with respect to the disc are
expected to be very rare \citep[][hereafter
  MA+10]{MendezAbreu2010}. To date NGC~4698 \citep{Bertola1999},
NGC~4672 \citep{Sarzi2000}, and UGC~10043 \citep{Matthews2004} are the
only spiral galaxies known to host a prominent bulge sticking out from
the plane of the disc.

We decided to revisit the case of NGC~4698 since such a rare galaxy
represents an excellent test case to derive the intrinsic shape of the
bulge by applying the method of MA+10.
NGC~4698 is an Sab(s) spiral \citep[][hereafter RC3]{RC3} in the Virgo
Cluster. Its total $B$-band magnitude is $B_T=11.46$ mag (RC3), which
corresponds to $M_B = -19.69$ mag assuming a distance of 17 Mpc
\citep{Freedman1994}.
The inner region of the large elliptical-like bulge is elongated
perpendicular to the major axis of the main disc, i.e. it is a polar
bulge; this is also true for the outer parts of the bulge if a
parametric photometric decomposition is adopted
\citep{Bertola1999}. At the same time, the stellar \citep{Bertola1999,
  Corsini1999} and ionised-gas \citep{Bertola2000} components are
characterised by an inner velocity gradient and a central
zero-velocity plateau along the minor and major axes of the disc,
respectively. This kinematically-decoupled core is rotating
perpendicularly with respect to the galaxy main disc. It corresponds
to a nuclear stellar disc \citep[hereafter NSD;][]{Pizzella2002}.
The geometrical and kinematical orthogonal decoupling of NGC~4698 can
hardly be explained without invoking the acquisition of external
material from the galaxy outskirts \citep[see][]{Bertola2000}.

In this Letter, we improve the previous results by quantitatively
constraining the intrinsic shape of the bulge of NGC~4698 in
Section~\ref{sec:bulge}.  The analysis of a near-infrared (NIR) image
from the United Kingdom Infrared Telescope (UKIRT) allows us to map
the mass distribution and minimise the contamination from dust. We get
a more reliable estimate of the galaxy's structural parameters with
respect to \citet{Bertola1999} since they folded the eastern side of
their optical image around the galaxy major axis to deal with the
strongest dust lanes of the main disc. In addition, they assumed the
galaxy surface brightness to be the sum of a de Vaucouleurs bulge and
an exponential disc {\em with} orthogonal major axes. Here, no {\em a
  priori\/} choice is made about the orientation of the bulge.
The structure of the NSD detected by \citet{Pizzella2002} is analysed
in Section~\ref{sec:nsd} in greater detail. The photometric
decomposition of multi-band optical images of the galaxy nucleus
obtained with the {\em Hubble Space Telescope (HST)\/} allows us to
investigate the formation process of the NSD by studying its stellar
populations.
Our findings are discussed in Section~\ref{sec:conclusions}.

\section{The bulge}
\label{sec:bulge}

\subsection{United Kingdom Infrared Telescope imaging}

A $K-$band image of NGC~4698 is available in the first data release of
the Large Area Survey of UKIRT Infrared Deep Sky Survey (UKIDSS,
\citealt{Lawrence2007}). It was obtained on 23 January 2007 with a
total exposure time of 40 s.
The 3.8-m UKIRT telescope is operated in Mauna Kea Observatory
(Hawaii, USA). It mounted the Wide Field Camera (WFCAM) with four
Rockwell Hawaii-II devices. Each of them consists of $2048\times2048$
HgCdTe detectors of $18 \times 18$ $\mu$m$^2$ and covers a field of
view of $13.7 \times 13.7$ arcmin$^2$ with an image scale of $0.4$
arcsec pixel$^{-1}$. The gain and read-out noise were 4.5 $e^-$
count$^{-1}$ and 25 $e^-$ rms, respectively.
The image was reduced, cleaned for cosmic rays, sky subtracted, and
flux calibrated with the WFCAM pipeline \citep{Hambly2008}. A
two-dimensional fit with a circular Gaussian to the field stars in the
resulting image yielded a $\rm FWHM = 0.76$ arcsec.

The photometric decomposition of the WFCAM image of NGC~4698 was
performed using the Galaxy Surface Photometry Two-Dimensional
Decomposition ({\sc GASP2D}) algorithm by MA+08, which yields the
structural parameters for a S\'ersic bulge and an exponential main
disc.

\begin{figure}
\psfig{file=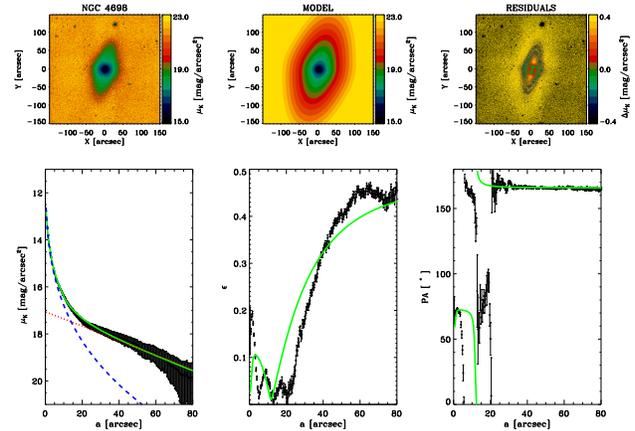,width=8.5cm,angle=0,clip=}
\caption{Two-dimensional photometric decomposition of NGC~4698. Top
  row: UKIDSS $K$-band image (left-hand panel), best-fit image
  (central panel), and residual (i.e., observed-model) image
  (right-hand panel).  North is up and East is left. Bottom row:
  Ellipse-averaged radial profiles of the surface brightness
  (left-hand panel), ellipticity (central panel), and position angle
  (right-hand panel) measured in the UKIDSS (dots) and model image
  (green continuous line).  The dashed blue and dotted red lines
  represent the intrinsic surface-brightness radial profiles of the
  bulge and main disc, respectively.}
\label{fig:UKIDSS}
\end{figure}

The fitting algorithm relies on a $\chi^2$ minimisation of the
intensities in counts, for which we must adopt initial trial
parameters that are as close as possible to their final values.  The
trial values of the effective surface brightness, effective radius,
shape parameter, position angle, and axial ratio of the bulge ($I_{\rm
  e}$, $r_{\rm e}$, $n$, PA$_{\rm b}$, and $q_{\rm b}$) and of the
central surface brightness, scalelength, position angle, and axial
ratio of the main disc ($I_0$, $h$, PA$_{\rm d}$, and $q_{\rm d}$)
were estimated from the analysis of the ellipse-averaged radial
profiles of surface brightness $\mu_K$, ellipticity $\epsilon$, and
position angle PA. The latter were measured using the {\sc IRAF} task
{\sc ELLIPSE} and analysed by following the prescriptions of MA+08.
Starting from these initial trial parameters the different photometric
models of the surface brightness were fitted iteratively to the galaxy
image. Each image pixel was weighted according to the variance of its
total observed photon counts due to the contribution of both galaxy
and sky, and determined assuming photon noise limitation and taking
the detector read-out noise into account.  Seeing effects were also
taken into account by convolving the model image with a circular
Gaussian point spread function (PSF) with a FWHM matching the observed
one. The convolution was performed as a product in Fourier domain
before the least-squares minimisation. Only the image pixels with an
intensity larger than 0.5 times the sky standard deviation were
included in the fit. Foreground stars were masked and excluded from
the fit.
The result of the photometric decomposition is shown in
Fig.~\ref{fig:UKIDSS}.
The residual image shows the tightly-wound spiral arms aligned with
the disc major axis. Nevertheless, the fit was satisfactory. The model
surface brightness is consistent with the data within the error
bars. The differences of ellipticity and position angle between the
observed and model isophotes are smaller than 0.1 and $3^\circ$,
respectively. The PA is scattered in the radial range between about 5
and 20 arcsec where the isophotes are nearly round.

The errors on the fitted parameters were estimated through Monte Carlo
simulations on images of artificial galaxies. A set of disc galaxies
with a S\'ersic bulge and an exponential main disc with $7 < K_T < 8$
mag to bracket the magnitude of NGC~4698 ($K_T = 7.54$ mag,
\citealt{Kassin2006}) was generated. The artificial galaxies were
assumed to be observed at the distance of the Virgo Cluster taking
into account resolution effects. The parameters of the artificial
galaxies were randomly chosen in the ranges observed for nearby
S0/a-Sb galaxies by \citet{Mollenhoff2001}. Finally, a background
level and photon noise were added to the simulated images in order to
mimic the instrumental setup and signal-to-noise of the WFCAM image.
The relative errors in the fitted parameters of the artificial
galaxies were estimated by comparing the input and output values and
were assumed to be normally distributed. The standard deviation was
adopted as the $1\sigma$ error in the relevant parameter for the
bulge-disc decomposition.
The best-fitting values and their $3\sigma$ errors are $\mu_{\rm e} =
17.07 \pm 0.64$ mag arcsec$^{-2}$, $r_{\rm e} = 11.5 \pm 6.0$ arcsec,
$n = 3.46 \pm 0.42$, PA$_{\rm b} = 72\fdg8 \pm 0\fdg9$, and $q_{\rm b}
= 0.88 \pm 0.03$ for the bulge, and $\mu_0 = 17.06 \pm 0.72$ mag
arcsec$^{-2}$, $h = 35.4 \pm 18.0$ arcsec, PA$_{\rm d} = 165\fdg7 \pm
6\fdg6$, and $q_{\rm d} = 0.47 \pm 0.06$ for the main disc.

Systematic errors given by a wrong estimation of the PSF FWHM and sky
level are the most significant contributors to the error budget, since
the spiral arms do not affect the result and no further component is
observed. We estimated a $3\sigma$ error of 2 and 1 per cent for the
sky level and PSF FWHM, respectively. We analysed the artificial
galaxies by adopting the correct sky level and a PSF FWHM that was 2
per cent larger (or smaller) than the actual one or the correct PSF
FWHM and a sky level that was 1 per cent larger (or smaller) than the
actual one. To derive the intrinsic shape of the bulge we are
interested in $q_{\rm b}$, $q_{\rm d}$, PA$_{\rm b}$, and PA$_{\rm
  d}$. Their $3\sigma$ errors including systematics are smaller than
15 per cent.

\subsection{Intrinsic shape of the bulge}

The method developed by MA+10 was applied to derive the intrinsic
shape of the bulge of NGC~4698. It is based upon the geometrical
relationships between the projected and intrinsic shapes of the bulge
and its surrounding disc. The bulge is assumed to be a triaxial
ellipsoid with semi-axes of length $A$ and $B$ in the equatorial plane
and $C$ along the polar axis. The bulge shares the same centre and
polar axis as its main disc, which is circular and lies on the
equatorial plane of the bulge. The intrinsic shape of the bulge is
recovered from the bulge ellipticity parameter $e = (1-q_{\rm
  b}^2)/(1+q_{\rm b}^2) = 0.13$, the twist angle $\rm \delta =
180^\circ-|PA_{\rm b} -PA_{\rm d}| = 87\fdg1$ between the major axes
of the bulge and main disc (i.e., the line of nodes), and the main
disc inclination $\theta = \arccos{q_{\rm d}} = 61\fdg7$.

\begin{figure}
\vspace{-0.5cm} \psfig{file=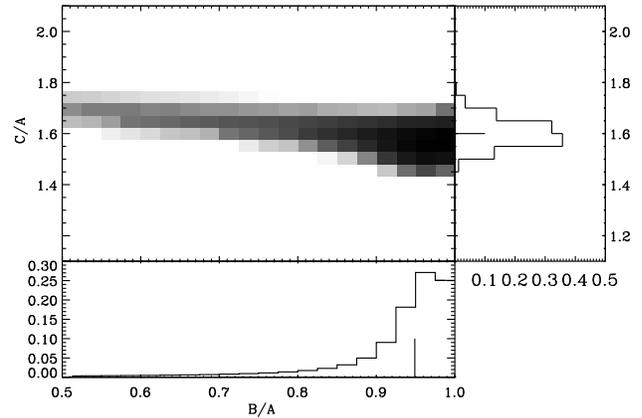,width=8.7cm, clip=}
  \caption{The distribution of the axial ratios of the bulge of
    NGC~4698. The median value of $B/A$ and $C/A$ is marked with a
    segment in the bottom and right-hand panel, respectively.}
  \label{fig:shape}
\end{figure}

The relation between the intrinsic and projected parameters depends
only on the spatial position of the bulge, i.e., on the angle $0 \le
\phi \le 90^\circ$ measured on the bulge equatorial plane between the
principal axis corresponding to $A$ and the line of nodes.  The
equatorial ellipticity $Z=B^2/A^2$ and intrinsic flattening
$F=C^2/A^2$ of the bulge are given by:

\begin{eqnarray}
\lefteqn{ \frac{\sin{\left(2\phi_C\right)}\,\sin^2{\theta}}{\cos^2{\theta}} F =  
 - \sin{\phi_B}\cos{\left(2\phi_C - \phi_B\right)} 
 \left(1+Z\right)^2 \nonumber } \\  
&+   \sin{\left(2\phi_C-\phi_B\right)}
\sqrt{\left(1-Z\right)^2 - \sin^2{\phi_B}\left(1+Z\right)^2},
\label{eqn:vartie}
\end{eqnarray}
%
where:

\begin{eqnarray}
\phi_B & = & \arctan{\frac{e\,\sin{2\delta}}{\cos{\theta}\,\left( 1+e\,\cos{2\delta}\right)}} \;= \;1\fdg8, \\
\phi_C & = & \frac{1}{2} \arctan{\frac{2\,e\,\sin{2\delta}\,\cos{\theta}}{e\,\cos{2\delta}\,\left(1+\cos^2{\theta}\right)-\sin^2{\theta}}} \;= \;89\fdg6.
\label{eqn:phi}
\end{eqnarray}
%
%
Since the axial ratios $B/A$ and $C/A$ are both functions of the same
variable $\phi$, their probabilities are identical, i.e., for a given
value of $B/A$ with probability $P(B/A)$, the corresponding value of
$C/A$ obtained by Eq. \ref{eqn:vartie} has a probability
$P(C/A)=P(B/A)$.
This allows to obtain the range of possible values of $B/A$ and $C/A$
for the bulge of NGC~4698 and to constrain its most probable intrinsic
shape by adopting the probabilities $P(Z)$ and $P(F)$ derived by
following MA+10.
Fig. \ref{fig:shape} shows the distribution of $B/A$ and $C/A$
calculated via Monte Carlo simulations. We randomly chose 5000
geometric configurations assuming a Gaussian distribution of $q_{\rm
  b}$, $q_{\rm d}$, PA$_{\rm b}$, and PA$_{\rm d}$ taking into account
their $3\sigma$ errors. For each geometric configuration we derived
1000 values of $B/A$ and $C/A$ according to their probability
distribution functions (see MA+10). Therefore, the resulting median
values of the axial ratios and the $3\sigma$ confidence intervals from
their cumulative distribution are $B/A=0.95^{+0.05}_{-0.91}$ and
$C/A=1.60^{+0.18}_{-0.12}$, respectively. The bulge of NGC~4698 is
elongated orthogonal to the main disc and the probability that it has
an equatorial section ($B/A<0.95$) is 50 per cent.

\section{The nuclear stellar disc}
\label{sec:nsd}

\subsection{Hubble Space Telescope imaging}

The Wide Field Planetary Camera 2 (WFPC2) images of NGC~4698 obtained
with the filters {\em F450W\/} (Prop. Id. 9042, P.I. S. J. Smartt),
{\em F606W\/} (Prop. Id. 6359, P.I. M. Stiavelli), and {\em F814W\/}
(Prop. Id. 9042, P.I. S. J. Smartt) were retrieved from the {\em
  HST\/} Science Data Archive.
The {\em F606W\/} images were taken by centring the galaxy nucleus on
the Planetary Camera (PC), whereas the others were taken with the Wide
Field Camera (WFC). The PC and WFC detectors are Loral CCDs with $800
\times 800$ pixels and a pixel size of $15 \times 15$ $\mu$m$^2$.  The
PC image scale of $0.046$ arcsec pixel$^{-1}$ yields a field of view
of about $36\times 36$ arcsec$^2$. Each WFC detector covers $72\times
72$ arcsec$^2$ with $0.091$ arcsec pixel$^{-1}$.
To help in identifying and correcting cosmic ray events, different
exposures were taken with each filter. The total exposure times was
460 s for the {\em F450W\/} and {\em F814W\/} filters and 600 s for
the {\em F606W\/} filter.  The telescope was always guided in fine
lock, giving a typical rms tracking error per exposure of $0.005$
arcsec.
The images were reduced using the {\sc CalWFPC} reduction pipeline in
{\sc IRAF} \citep{McMaster2008}. Subsequent analysis including
alignment and combination, rejection of cosmic rays, and sky
subtraction was performed using {\sc IRAF} standard tasks.
The flux calibration to the Vega magnitude system was performed
following \citet{Holtzman1995}.
The flux calibration to the Vega magnitude system in the {\em HST\/}
passbands was performed following \citet{Holtzman1995}.

\begin{figure}
\psfig{file=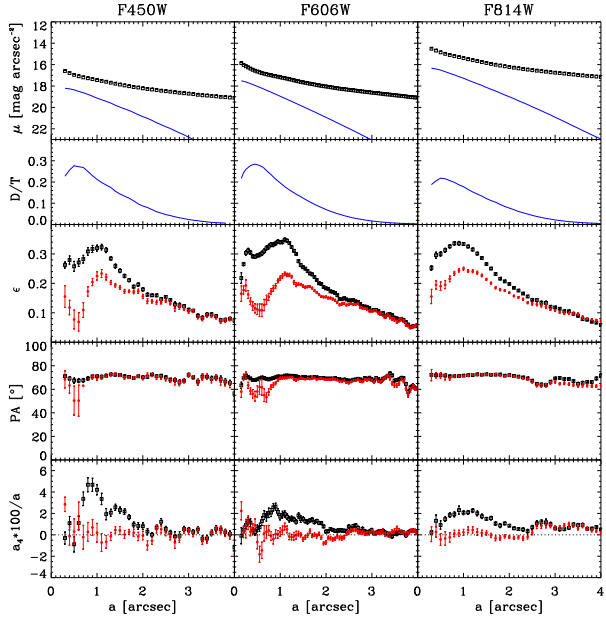,width=8.7cm,angle=0, clip=}
\caption{Isophotal parameters of the nuclear region of NGC~4698 as a
  function of the isophotal semi-major axis based on the analysis of
  the surface-brightness distribution measured in the {\em F450W\/}
  (left-hand panels), {\em F606W\/} (central panels), and {\em
    F814W\/} (right-hand panels) images, respectively. From top to
  bottom: Surface-brightness radial profiles of the galaxy (open black
  squares) and NSD (after convolution with the {\em HST\/} PSF, solid
  blue line). Radial profile of the NSD-to-total luminosity
  ratio. Radial profiles of the galaxy ellipticity, position angle,
  and fourth cosine Fourier coefficient before (open black squares)
  and after (filled red circles) the subtraction of the best-fitting
  model for the NSD.}
  \label{fig:HST}
\end{figure}

Following \citet{Pizzella2002} the unsharp-masked image of each WFPC2
frame was built to first gauge the structure and extent of the NSD in
all the available images.
This procedure enhanced any surface-brightness fluctuation and
non-circular structure extending over a spatial region comparable to
the standard deviation of the smoothing Gaussian.
The galaxy nucleus clearly reveals the presence of a highly elongated
structure in all the images. This is the NSD and its location,
orientation, and size are the same in all the observed
passbands. This nuclear structure is associated with a central
increase in ellipticity, $\epsilon$, and fourth cosine Fourier
coefficient, $a_4$, as measured by performing an isophotal analysis
using {\sc ELLIPSE} and shown in Fig.~\ref{fig:HST}.

The photometric parameters of the NSD were derived in the different
passbands using the method by \citet[][hereafter SB95]{Scorza1995} as
implemented by \citet{Morelli2004} although with a different treatment
of the PSF.  The photometric decomposition was performed independently
for each band-pass.
The SB95 method is based on the assumption that the isophotal
disciness is the result of the superimposition of a host spheroidal
component and an inclined exponential disc. Both the components are
assumed to have perfectly elliptical isophotes with constant but
different ellipticity. The method consists of the iterative
subtraction of an infinitesimally thin disc model characterised by an
exponential surface-brightness profile with central surface brightness
$I_{0, {\rm NSD}}$, scalelength $h_{\rm NSD}$, axial ratio $q_{\rm
  NSD}$, and position angle PA$_{\rm NSD}$.  The NSD parameters are
adjusted until the departures from perfect ellipses are minimised
(i.e., $a_{4}\approx0$).
To properly derive the photometric parameters of the NSD, it is
important to account for the {\em HST\/} PSF. For each nuclear disc
model, the disc-free image of the galaxy was obtained from the galaxy
image by subtracting the nuclear disc model after convolving with the
{\em HST\/} PSF.  The adopted PSF model was calculated with the
TINYTIM package taking into account the instrumental setup and
position of the NSD on the given image \citep{Krist1999}.
The best-fitting values of the NSD parameters and their $3\sigma$
errors were derived as in \citet{Morelli2004} and are listed in
Tab.~\ref{tab:disc_parameters}. The NSD inclination is calculated as
$i_{\rm NSD} =\arccos{q_{\rm NSD}}$. The NSD is elongated like the
galaxy bulge as is evident from its position angle.
The comparison between the isophotal parameters of NGC~4698 measured
before and after the subtraction of the best-fitting model of the NSD
are shown in Fig.~\ref{fig:HST}.  

\citet{Pizzella2002} had already analysed the {\em F606W\/} image with
the SB95 method. They deconvolved the galaxy image from the effects of
the PSF using the Richardson-Lucy method.
We decided to repeat the photometric decomposition since both an
homogeneous photometric decomposition and an estimate of the errors on
the fitted parameters are required to properly compare the structural
parameters of the NSD in the different passbands. Moreover, this
allowed us to verify that the NSD parameters obtained by subtracting
the PSF-convolved disc model from the galaxy image are in agreement
within the errors with those obtained by subtracting the unconvolved
disc model from a deconvolved galaxy image. To this aim the conversion
to Johnson $V$ band has to be taken into account for a proper
comparison of $I_{0,{\rm NSD}}$ and was calculated with {\sc IRAF}
task {\sc SYNPHOT} ($V-{\it F606W} = 0.27$ for a spiral galaxy).

\begin{table}
\caption{Photometric parameters of the nuclear stellar disc.} 
\begin{tabular}{lccccc}
\hline
\multicolumn{1}{c}{Filter} &
\multicolumn{1}{c}{$I_{0,{\rm NSD}}$} &
\multicolumn{1}{c}{$h_{\rm NSD}$} &
\multicolumn{1}{c}{$i_{\rm NSD}$} &
\multicolumn{1}{c}{PA$_{\rm NSD}$} \\
\multicolumn{1}{c}{} &
\multicolumn{1}{c}{[mag arcsec$^{-2}$]} &
\multicolumn{1}{c}{[pc]} &
\multicolumn{1}{c}{[$^\circ$]} &
\multicolumn{1}{c}{[$^\circ$]} \\

\hline
{\em F450W} & $17.39^{+0.24}_{-0.04}$ & $50.4^{+1.8}_{-6.7}$  & $79.4^{+3.3}_{-6.4}$ & $70\pm2$ 
\\
{\em F606W} & $16.97^{+0.49}_{-0.12}$ & $45.2^{+4.1}_{-6.0}$  & $73.4^{+2.2}_{-4.7}$ & $71\pm2$ \\
{\em F814W\/} & $15.63^{+0.45}_{-0.29}$ & $50.4^{+9.8}_{-11.1}$ & $75.5^{+9.0}_{-5.2}$ & $68\pm2$ \\
\hline
\end{tabular}
\label{tab:disc_parameters}
\end{table}

\subsection{Formation of the nuclear stellar disc}

The structural parameters of the NSD (i.e., scalelength, inclination,
and position angle) are constant within the errors in all the
available WFPC2 images. Therefore, the location, orientation, and size
of the NSD do not depend on the observed passband. This also implies
the absence of colour gradients in the NSD, an important constraint on
the star formation process.
The mean values of the NSD parameters are $\langle h_{\rm NSD} \rangle
= 48.7$ pc, $\langle i_{\rm NSD} \rangle = 76\fdg1$, and $\rm \langle
PA_{\rm NSD} \rangle = 69\fdg7$.
The size and luminosity are consistent with those of the other NSDs
detected so far \citep[see][for a census]{Ledo2010}.
The NSD is oriented as is the bulge, and its major axis is
perpendicular to that of the main disc, i.e. it is a polar NSD.

A linear combination of single-age stellar population synthesis models
was used by \citet{Sarzi2005} to interpret an {\em HST\/} spectrum
measured within the central 0.13 arcsec (11 pc) of NGC~4698. The
continuum spectral energy distribution between 3000-5700 \AA\ was best
fitted by a very old stellar population (10 Gyr) with no evidence for
stars younger than 5 Gyr in the galaxy nucleus, where the NSD light
contamination is large ($\sim20$ per cent, Fig. \ref{fig:HST}). The
light fraction from stars younger than 1 Gyr rises to 2 per cent if
supersolar-metallicity models are considered. This implies that the
centre of the NSD is made mostly of old stars.
Models for chemical and spectro-photometric evolution of galaxy discs
predict a strong evolution with age of their colour profiles as direct
consequence of an inside-out formation \citep{Boissier1999,
  Prantzos2000}. In this scenario colour gradients are produced early
on in the disc centre and propagate outwards. The maximum colour
gradient measured for the NSD ($|d(B-R)/dr| < 0.05$ mag kpc$^{-1}$)
falls short of the model predictions and it is not consistent with a
picture wherein the nuclear disc assembled inside out in the last 5-10
Gyr.

\section{Discussion and conclusions}
\label{sec:conclusions}

The structure of the nucleus and bulge of the Virgo spiral galaxy
NGC~4698 was investigated through a detailed analysis of {\em HST\/}
optical and UKIRT near-infrared images, respectively.
The galaxy is known to host a nuclear disc of gas and stars which is
rotating perpendicularly with respect to the galaxy main
disc. Moreover, the bulge and main disc appear on the sky elongated
perpendicular to each other \citep{Bertola1999, Bertola2000,
  Pizzella2002}.

The equatorial ellipticity and intrinsic flattening of the polar bulge
were obtained following MA+10 from the apparent ellipticity of the
bulge, twist angle between the bulge and main disc, and main disc
inclination measured in the UKIDSS near-infrared image by adopting the
photometric decomposition method by MA+08.
The bulge of NGC~4698 is remarkably elongated in a perpendicular
direction with respect to the disc plane
($C/A=1.60^{+0.18}_{-0.12}$). Although the consistency with
axisymmetry was recognised ($B/A=0.95^{+0.05}_{-0.91}$), still the
probability that it is significantly triaxial (i.e., with an
equatorial elliptical section with $B/A<0.95$) is 50 per cent.
The central surface brightness, scalelength, inclination, and
position angle of the polar NSD in all the available {\em HST\/}
images were measured by assuming it is an infinitesimally thin
exponential disc and applying the photometric decomposition method of
SB95 as implemented by \citet{Morelli2004}. The size, orientation, and
location of the polar NSD do not depend on the observed passband, as
already observed for the few other NSDs for which a detailed
multi-band photometric analysis was performed
(\citealt{Krajnovic2004}, \citealt{Morelli2010}).

The combination of these new results about the complex structure of
NGC~4698 gives us the opportunity to gain further insight on the
formation of NSDs with the goal of motivating numerical modelling to
test and refine this scenario. The kinematical decoupling between two
components of a galaxy suggests the occurrence of an accretion event
or merging \citep{BertolaCorsini1999}. Therefore, it is
straightforward to explain the existence of the orthogonally-rotating
dynamically-cold nuclear disc in NGC~4698 as the end result of the
acquisition of external gas by the pre-existing galaxy. Gas
dissipation is indeed a necessary ingredient since purely stellar
dynamical mergers can not form a nuclear disc \citep{Hartmann2011}.
In NGC~4698 the accreted gas settled on the principal plane
perpendicular to the shortest axis of the triaxial bulge (i.e.,
perpendicular to the galaxy main disc) and formed stars. The major
axis of the NSD is elongated in the same direction as the bulge.
The link between the angular momentum transport from galactic to
nuclear scales and formation of NSDs has been recently investigated by
\citet{Hopkins2010}. 
The stellar population in the centre of NGC~4698 is very old
\citep{Sarzi2005}. The absence of colour gradients in the NSD is
explained if either the star formation homogeneously occurred all over
the extension of the disc or the NSD assembled through an inside-out
process that ended more than 5 Gyr ago.

\section*{Acknowledgments}
We are grateful to Victor P. Debattista for valuable comments. This
work is supported by Padua University (grant 60A02-1283/10) and
Italian Space Agency (contract ASI-INAF I/009/10/0). LM acknowledges
support from Padua University (grant CPS0204).  JMA is partially
funded by the Spanish MICINN (Consolider-Ingenio 2010 Program grant
CSD2006-00070 and grants AYA2007-67965-C03-01 and
AYA2010-21887-C04-04).

\label{lastpage}

\end{document}